\documentclass[11pt,a4paper,hyperref]{article}

\usepackage{amsmath}
\usepackage{amssymb}
\usepackage{amsfonts}
\usepackage{slashed}
\usepackage{mathrsfs}
\usepackage[breaklinks,colorlinks,citecolor=blue,urlcolor=blue,linkcolor=blue]{hyperref}
\usepackage[numbers,sort&compress]{natbib}
\usepackage{graphicx}

\begin{document}
%

%
\title{Possible large $CP$ violation in three body decays of heavy baryon}
%
%
%
\author{ Zhen-Hua~Zhang$^{1}$, Chao~Wang$^{2}$, and Xin-Heng~Guo$^{2}$
\\ $^{1}$School of Nuclear Science and Technology, University of South China, \\ Hengyang, Hunan 421001, China.
\\$^{2}$ College of Nuclear Science and Technology, Beijing Normal University, \\Beijing 100875, China.}

%
\maketitle
%
%

%
\begin{abstract}
We propose a new mechanism which can introduce large $CP$ asymmetries in the phase spaces of three-body decays of heavy baryons.
In this mechanism, a large $CP$ asymmetry is induced  by the interference of two intermediate resonances, which subsequently decay into two different combinations of final particles.
We apply this mechanism to the decay channel $\Lambda_b^0 \to p \pi^0\pi^-$, and find that the differential $CP$ asymmetry can reach as large as $50\%$, while the regional $CP$ asymmetry can reach as large as $16\%$ in the interference region of the phase space.
\end{abstract}
%

\section{Introduction}
%
$CP$ violation is an important phenomenon in particle physics.
Although it has been discovered in the mixing and decay processes of $K$ and $B$ meson systems, including the first discovery of $CP$ violation in $K$ system \cite{Christenson:1964fg}, no $CP$ violation was established in the baryon sector, except an evidence in the decay channel $\Lambda_b^0\to p K^-$ \cite{Aaltonen:2011qt}.
Within the Standard Model, $CP$ violation is originated from the weak phase in the Cabibbo--Kobayashi--Maskawa (CKM) matrix \cite{Kobayashi:1973fv}, along with a strong phase which usually arises from strong interactions.
One reason for the smallness of $CP$ violation is that the strong phases are usually small, especially when the strong phases come from a scale that is much larger than the QCD scale. 
However, non-perturbative effects of the strong interaction at low scales provide possibilities for large strong phases, and hence, large $CP$ violation.

Three-body decays of heavy hadrons can be dominated by intermediate resonances in certain regions of the phase space.
When two resonances decay into two different combinations of final particles, it is possible for them to dominate in the same region of the phase space.
As a result, the interference effect together with a possible large strong phase can generate a large $CP$ asymmetry. 

%
\section{Differential $CP$ asymmetry}
%
It gets more interesting when one applies the aforementioned interference effect to the decay process of heavy baryons. 
For the decay process $\Lambda_b^0\to p \pi^0\pi^-$, there is an overlap region in the phase space for resonances $\rho^-(770)$ and $N^+(1440)$, which lies right in the corner of the phase space.
The decay amplitude for $\Lambda_b^0\to p \pi^0\pi^-$ can be expressed as
\begin{eqnarray}\label{DecayAmplitude}
\mathscr{M}=\frac{\langle p \pi^0|\hat{\mathscr{H}}_1|N^+\rangle \langle \pi^- N^+|\hat{\mathscr{H}}_{\rm{eff}}|\Lambda_b^0\rangle
}{s_0-m_{N}^2+i m_{N}\Gamma_{N}}+
\frac{\langle\pi^0\pi^-|\hat{\mathscr{H}}_2|\rho^-\rangle \langle p \rho^-|\hat{\mathscr{H}}_{\rm{eff}}|\Lambda_b^0\rangle
}{s-m_{\rho}^2+i m_{\rho}\Gamma_{\rho}},
\end{eqnarray}
in the overlap region of the phase space, where $\hat{\mathscr{H}}_{\rm{eff}}$ is the effective Hamiltonian for the weak decays, 
$\hat{\mathscr{H}}_1$ and $\hat{\mathscr{H}}_2$ are the formal Hamiltonian for the strong decays in which the magnitudes of the coupling constants can be determined from experiments, $s$ and $s_0$ are the invariant mass squares of the systems $\pi^0\pi^-$ and $p \pi^0$, respectively, $m_\rho$, $m_N$, $\Gamma_\rho$, and $\Gamma_N$ are the masses and decay widths of $\rho^0(770)$ and $N^+(1440)$, respectively, and the summation over the polarizations of the intermediate particles is understood.
The effective Hamiltonian $\hat{\mathscr{H}}_{\rm{eff}}$ takes the form \cite{Buchalla:1995vs}
\begin{eqnarray}\label{EffectiveHamiltonian}
\hat{\mathscr{H}}_{\rm{eff}}&=&\frac{G_F}{\sqrt{2}}\left[V_{ub}V_{ud}^*(c_1O_1^u+c_2O_2^u)+V_{cb}V_{cd}^*(c_1O_1^c+c_2O_2^c)-V_{tb}V_{td}^*\sum_{i=3}^{10}c_iO_i\right]\nonumber\\
&&+h.c.,
\end{eqnarray}
where $G_F$ is the Fermi constant, $V_{ud}$, $V_{ub}$, $V_{cd}$, $V_{cb}$, $V_{td}$, and $V_{tb}$ are the CKM matrix elements, $c_i$ $(c_i=1,\cdots, 10)$ is the Wilson constant, and $O_i$ is the four-Fermion operator, which takes the form
\begin{equation}\label{4FerminonOperator}
\begin{aligned}
O_1^q =&{} \bar{d}_\alpha\gamma_\mu(1-\gamma_5)q_\beta\bar{q}_\beta\gamma^\mu(1-\gamma_5)b_\alpha,\\
O_2^q =&{} \bar{d}\gamma_\mu(1-\gamma_5)q\bar{q}\gamma^\mu(1-\gamma_5)b,\\
O_3=&{} \bar{d}\gamma_\mu(1-\gamma_5)b\sum_{q'}\bar{q'}\gamma^\mu(1-\gamma_5)q',\\
O_4 = &{} \bar{d}_\alpha\gamma_\mu(1-\gamma_5)b_\beta \sum_{q'}\bar{q}'_\beta\gamma^\mu(1-\gamma_5)q'_\alpha,\\
O_5 = &{} \bar{d}\gamma_\mu(1-\gamma_5)b\sum_{q'}\bar{q'}\gamma^\mu(1+\gamma_5)q',\\
O_6 = &{} \bar{d}_\alpha\gamma_\mu(1-\gamma_5)b_\beta \sum_{q'}\bar{q}'_\beta\gamma^\mu(1+\gamma_5)q'_\alpha,\\
O_7 =&{} \frac{3}{2}\bar{d}\gamma_\mu(1-\gamma_5)b\sum_{q'}e_{q'}\bar{q'}\gamma^\mu(1+\gamma_5)q',\\
O_8 = &{} \frac{3}{2}\bar{d}_\alpha\gamma_\mu(1-\gamma_5)b_\beta \sum_{q'}e_{q'}\bar{q}'_\beta\gamma^\mu(1+\gamma_5)q'_\alpha,\\
O_9 = &{} \frac{3}{2}\bar{d}\gamma_\mu(1-\gamma_5)b\sum_{q'}e_{q'}\bar{q'}\gamma^\mu(1-\gamma_5)q',\\
O_{10}= &{} \frac{3}{2}\bar{d}_\alpha\gamma_\mu(1-\gamma_5)b_\beta \sum_{q'}e_{q'}\bar{q}'_\beta\gamma^\mu(1-\gamma_5)q'_\alpha,\\
\end{aligned}
\end{equation}
with $d$, $b$, $q$, and $q'$ being quark fields and $\alpha$ and $\beta$ being colour indices.

Under the factorization hypothesis, the weak decay amplitudes can be expressed as
\begin{equation}\label{Amp1}
\langle \pi^-N^+|\mathscr{H}_{\rm eff}|\Lambda_b^0\rangle=i\eta^N\overline{u}_N\slashed{p}_{\pi^-}(1-\gamma_5)u_{\Lambda_b},
\end{equation}
\begin{equation}\label{Amp2}
\langle \rho^-p|\mathscr{H}_{\rm eff}|\Lambda_b^0\rangle=\eta^p m_\rho\overline{u}_N\slashed{\epsilon}_{\rho^-}(1-\gamma_5)u_{\Lambda_b},
\end{equation}
where $\epsilon_{\rho^-}$ is the polarization vector of $\rho^-$, $u_{N}$ and $u_{\Lambda_b}$ are the spinors for $N^+(1440)$ and $\Lambda_b$, respectively,
\begin{equation}
\eta^N=\frac{G_F}{\sqrt{2}}f_\pi F^{\Lambda_b\to N^+}\left\{a_2V_{ub}V_{ud}^*-V_{tb}V_{td}^*\left[(a_4+a_{10})-\frac{2m_\pi^2(a_6+a_8)}{(m_u+m_d)m_b}\right]\right\},
\end{equation}
\begin{equation}
\eta^p=\frac{G_F}{\sqrt{2}}f_\rho F^{\Lambda_b\to p}\left\{a_2V_{ub}V_{ud}^*-V_{tb}V_{td}^*\left[(a_4+a_{10})\right]\right\},
\end{equation}
with $f_\pi$ being the decay constant of the pion, $F^{\Lambda_b\to N^+}$ and $F^{\Lambda_b\to p}$ being the form factors for the transition $\Lambda_b\to N^+(1440)$ and $\Lambda_b\to p$, respectively, and $a_i=c_i+c_{i-1}/N_c$ for even $i$.

Because of the non-purterbative effects of strong interactions, there can be a relative strong phase between the coupling constants of $\hat{\mathscr{H}}_1$ and $\hat{\mathscr{H}}_2$.
We will denote this relative phase by $\delta$ and treat it as a free parameter.
The strong decay amplitudes are then expressed as 
\begin{equation}
\langle p \pi^0|\hat{\mathscr{H}}_1|N^+\rangle=ig_1\overline{u}_p\gamma_5 u_N,
\end{equation}
and
\begin{equation}
\langle\pi^0\pi^-|\hat{\mathscr{H}}_2|\rho^-\rangle=e^{i\delta} g_2 (p_{\pi^-}-p_{\pi^0})\cdot \epsilon_{\rho^-},
\end{equation}
respectively, where the effective coupling constants $g_1$ and $g_2$ can be expressed as
\begin{eqnarray}
g_1^2
\!\!\!\!&=&\!\!\!\!
\frac{8\pi m_N^2\Gamma_{N^+\to N\pi}}{3\lambda_N(m_N^2+m_\rho^2-2m_Nm_p-m_\pi^2)},
\\
g_2^2
\!\!\!\!&=&\!\!\!\!
\frac{6\pi m_\rho^2\Gamma_{\rho^-\to \pi^0\pi^-}}{\lambda_\rho^3},
\end{eqnarray}
with $m_p$ being the mass of proton, $\Gamma_{N^+\to N\pi}$ and $\Gamma_{\rho^-\to \pi^0\pi^-}$ being the partial decay widths for $N^+(1440)\to N(939)\pi$ and $\rho^-(770)\to \pi^0\pi^-$, respectively, and
\begin{eqnarray}
\lambda_N
\!\!\!\!&=&\!\!\!\!
\frac{1}{2m_N}
\sqrt{\left[m_N^2-(m_p+m_\pi)^2\right]\cdot\left[m_N^2-(m_p-m_\pi)^2\right]},
\\
\lambda_\rho
\!\!\!\!&=&\!\!\!\!
\frac{1}{2}
\sqrt{m_\rho^2-4(m_\pi)^2}.
\end{eqnarray}

The differential $CP$ asymmetry is then defined as
\begin{equation}
A_{CP}=\frac{\overline{\left|\mathscr{M}\right|^2}-\overline{\left|\bar{\mathscr{M}}\right|^2}}{\overline{\left|\mathscr{M}\right|^2}+\overline{\left|\bar{\mathscr{M}}\right|^2}},
\end{equation}
where $\bar{\mathscr{M}}$ is the decay amplitude of the $CP$ conjugate process, $\overline{\Lambda_b^0}\to \overline{p}\pi^+\pi^0$, and the overlines above $\left|\mathscr{M}\right|^2$ and $\left|\bar{\mathscr{M}}\right|^2$ represent averaging and summing over the spin states of the initial and final particles, respectively.
After some algebra, one has 
\begin{eqnarray}\label{Msqured}
\overline{\left|\mathscr{M}\right|^2}
\!\!\!\!&=&\!\!\!\!
\Big\{|\lambda_{1}|^2\left[(m_{\Lambda_b}^2-s_-)(s_0-m_p^2)-m_\pi^2(m_{\Lambda_b}-m_p)^2+m_\pi^4\right]\nonumber\\
&&+|\lambda_{2}|^2\left[(m_{\Lambda_b}^2-s_0)(s_--m_p^2)-m_\pi^2(m_{\Lambda_b}-m_p)^2+m_\pi^4\right]\nonumber\\
&&+2\mathscr{R}\left(\lambda_{1}\lambda_{2}^*\right)\left[s_0s_-+m_{\Lambda_b}m_p(m_{\Lambda_b}^2-m_{\Lambda_b}m_p+m_p^2-s_0-s_-)-m_\pi^4\right]\!\!\Big\}\nonumber\\
&&+\Big\{m_{\Lambda_b}\to -m_{\Lambda_b}\Big\},
\end{eqnarray}
where $s_-$ is the invariant mass squared of the system $p\pi^-$,
\begin{eqnarray}
\label{lambda1}
\lambda_1
\!\!\!\!&=&\!\!\!\!
\frac{m_{\Lambda_b}^2-s_0}{m_{\Lambda_b}-m_p}\frac{g_1}{s_N}\eta^N+e^{i\delta}\frac{g_2}{s_\rho}m_\rho\eta^p,
\\
\label{lambda2}
\lambda_2
\!\!\!\!&=&\!\!\!\!
\frac{m_{\Lambda_b}(m_p-m_N)+m_pm_N-s_0}{m_{\Lambda_b}-m_p}\frac{g_1}{s_N}\eta^N-e^{i\delta}\frac{g_2}{s_\rho}m_\rho\eta^p,
\end{eqnarray}
and $s_N=s_0-m_N^2+im_N\Gamma_N$, $s_\rho=s-m_\rho^2+im_\rho\Gamma_\rho$.
In order to obtain the expression for $\overline{\left|\bar{\mathscr{M}}\right|^2}$, all one needs to do is to replace the CKM matrix elements in Eq. (\ref{Msqured}) with their complex conjugates.

In order to see where the $CP$ asymmetry arises, let's first display the weak and strong phases in $\lambda_1$ and $\lambda_2$ explicitly. 
For a fixed point in the phase space,  $\lambda_1$ and $\lambda_2$ can be expressed as
\begin{equation}
\label{lambdai}
\lambda_i=\lambda_i^{\text{Tree}}e^{i(\phi_i^{\text{Tree}}+\alpha_i^{\text{Tree}})}+\lambda_i^{\text{Penguin}}e^{i(\phi_i^{\text{Penguin}}+\alpha_i^{\text{Penguin}})},
\end{equation}
where $i=1,2$, $\lambda_i^{\text{Tree}}$ and $\lambda_i^{\text{Penguin}}$ are the tree and penguin parts of $\lambda_i$, respectively, $\phi_i^{\text{Tree}}$ and $\phi_i^{\text{Penguin}}$ are the corresponding weak phases, which take the values
\begin{equation}
\phi_i^{\text{Tree}}=\text{Arg}\left({V_{ub}V_{ud}^*}\right),
\end{equation}
and
\begin{equation}
\phi_i^{\text{Penguin}}=\text{Arg}\left({V_{tb}V_{td}^*}\right),
\end{equation}
$\alpha_i^{\text{Tree}}$ and $\alpha_i^{\text{Penguin}}$ are the corresponding strong phases, which originate mainly form the strong phase $\delta$ and the phases in the propagators. 
Since the strong phases are $CP$-even while the weak phases are $CP$-odd, with the aid of Eq. (\ref{Msqured}), it follows that the difference between $\overline{\left|\mathscr{M}\right|^2}$ and $\overline{\left|\bar{\mathscr{M}}\right|^2}$ takes the from
\begin{eqnarray}\label{DiffMsquare}
\overline{\left|\mathscr{M}\right|^2}-\overline{\left|\bar{\mathscr{M}}\right|^2}
\!\!\!\!&\sim&\!\!\!\! 
\sin\phi \Big[a\sin\left(\alpha_1^{\text{Penguin}}-\alpha_1^{\text{Tree}}\right)+b\sin\left(\alpha_2^{\text{Penguin}}-\alpha_2^{\text{Tree}}\right)
\nonumber\\
\!\!\!\!&&\!\!\!\! 
+c\sin\left(\alpha_1^{\text{Penguin}}-\alpha_2^{\text{Tree}}\right)+d\sin\left(\alpha_2^{\text{Penguin}}-\alpha_1^{\text{Tree}}\right)\Big],
\end{eqnarray}
where $\phi=\text{Arg}\left({V_{tb}V_{td}^*}/{V_{ub}V_{ud}^*}\right)$, $\alpha_i=\alpha_i^{\text{Penguin}}-\alpha_i^{\text{Tree}}$, and $a$, $b$, $c$, and $d$ are real quantities.
One can see from Eq. (\ref{DiffMsquare}) that both the strong and weak phases are  essential for $CP$ violation.
The difference in Eq. (\ref{DiffMsquare}) is proportional to the sine of the difference of the tree and penguin weak phases.
Besides, the four terms in Eq. (\ref{DiffMsquare}) are also proportional to the sine of the differences of the tree and penguin strong phases, respectively.

\begin{figure}
\includegraphics[width=\textwidth]{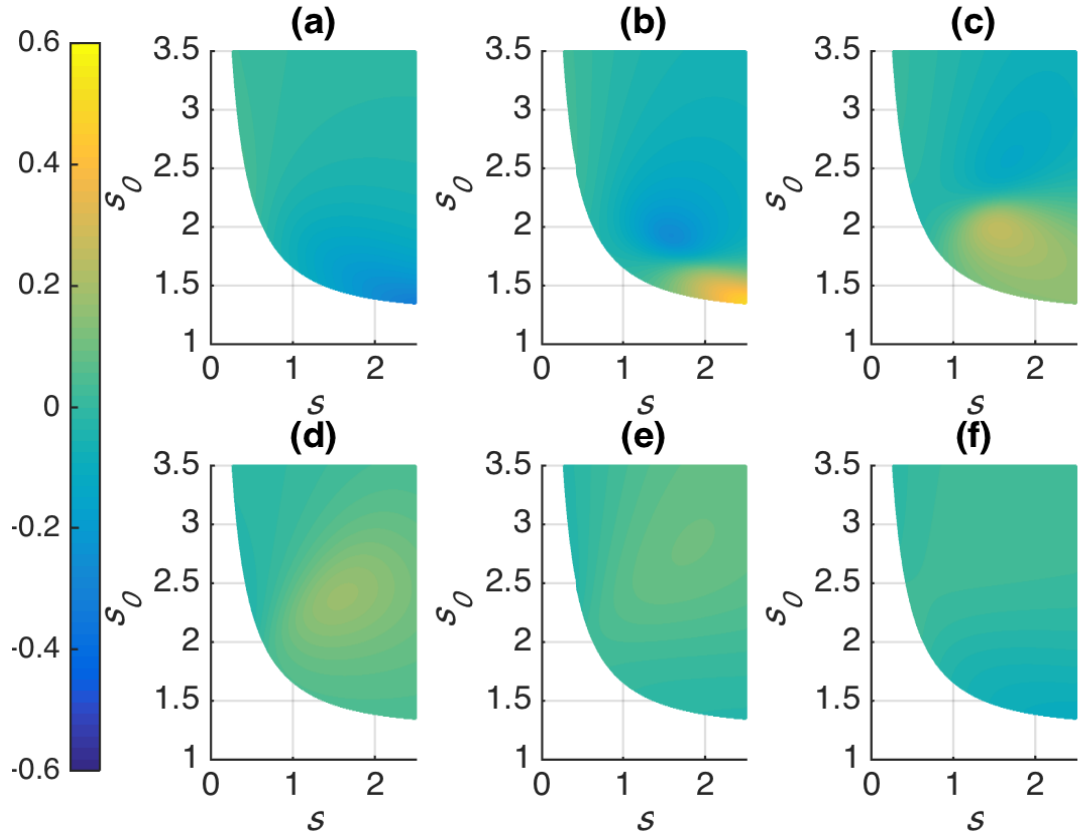}
\caption{\label{DifACP} Differential $CP$ asymmetry (in unit of $\%$) distributions in the overlap region of the phase space for various values of $\delta$. The six diagrams (a) to (f) correspond to $\delta$ taking values form 0  to $5\pi/3$ for every $\pi/3$. The invariant mass squares, $s$ and $s_0$, are in units of GeV$^2$.}
\end{figure}

In Fig. \ref{DifACP}, we present the differential $CP$ asymmetry distribution in the overlap region of the phase space for various values of $\delta$, from which one can see clearly that the interference effect of the two aforementioned resonances does result in a $CP$ asymmetry in the overlap region of the phase space.

Among the input parameters, the Wilson coefficients are taken from Ref. \cite{Buchalla:1995vs}, and the rest of the input parameters are from Particle Data Group \cite{Agashe:2014kda}.
In the heavy quark limit and for large recoil final light baryon, both of the from factors for $\Lambda_b\to p$ and $\Lambda_b\to N^+(1440)$ transitions reduce to a single form factor \cite{Mannel:2011xg}. 
Since the structure of $N^+(1440)$ is still not clear, the decay form factors for the transition  $\Lambda_b\to N^+(1440)$ is not available.
Therefore, we set in Fig. \ref{DifACP} the heavy-to-light baryonic tradition form factors equal to each other, i.e. $F^{\Lambda_b\to p}=F^{\Lambda_b\to N^+}$.

One interesting behaviour for the differential $CP$ asymmetry in Fig. \ref{DifACP} is that it can be as large as $50\%$ in the overlap region of the phase space.
Besides, one can see that the differential $CP$ asymmetry shows large anisotropic behaviour in the overlap region, especially when $s$ is away from the vicinity of $\rho^-(770)$ ($\sqrt{s}>m_\rho+\Gamma_\rho$).
The reason for this behaviour is that the amplitude of $\Lambda_b \to p\rho^-\to p\pi^0\pi^-$ is larger than that of  $\Lambda_b \to N^+(1440)\pi^-\to p\pi^0\pi^-$ when $s$ and $s_0$ are close to $m_{\rho^-}^2$ and $m_{N^+}^2$, respectively. 

%
\section{Regional $CP$ asymmetry}
%
 
In order to compare with future experiments, one has to consider the regional $CP$ asymmetry, which can be defined as
\begin{equation}
A_{CP}^{\Omega}=\frac{\Gamma^{\Omega}-\bar{\Gamma}^{\Omega}}{\Gamma^{\Omega}+\bar{\Gamma}^{\Omega}},
\end{equation}
where $\Omega$ is some region of the phase space, $\Gamma^{\Omega}$ and $\bar{\Gamma}^{\Omega}$ are the regional decay width for $\Lambda_b^0\to p \pi^0\pi^-$ and $\overline{\Lambda}_b^0\to \overline{p} \pi^0\pi^+$, respectively, with the former one taking the form
\begin{equation}
\Gamma^{\Omega}=\frac{1}{256\pi^3m_{\Lambda_b}^3}\int_{\Omega} dsds_{0} \overline{\left|\mathscr{M}\right|^2}.
\end{equation}

We will focus on an overlap region of the phase space which satisfies $m_\rho+\Gamma_\rho<\sqrt{s}< m_\rho+2\Gamma_\rho $, and $m_N-0.5\Gamma_N<\sqrt{s_0}< m_N+0.5\Gamma_N $, and denote it by $\Omega_{\rm{OL}}$.
The reason for chosing this region is of two folds.
First, we have to exclude the pollution of other resonances.
For the $\pi^0\pi^-$ system, one can easily check that $\rho^-(770)$ is the only dominated resonance for $\sqrt{s}<1.5$ GeV.
The amplitude of the first term in Eq. (\ref{DecayAmplitude}) still dominates even if $s$ is little bit away from the vicinity of $\rho^-(770)$.
For the $p\pi^-$ system, on the other hand, resonances such as $\Delta(1232)$, $N(1520)$, and $N(1535)$ could give comparable contributions besides $N^+(1440)$.
In order to exclude these resonances, we have to keep close to the vicinity of $N^+(1440)$.
Secondly, it is understandable that the contribution of the interference effect becomes more significant when the two amplitudes in Eq. (\ref{DecayAmplitude}) are comparable. 
One can easily check that the first term is much larger than the second one when both of the resonances $\rho^-(770)$ and $N^+(1440)$ are on the mass shell.
Consequently, we choose the region $\Omega_{\text{OL}}$ around the vicinity of $N^+(1440)$ but a bit further away from the vicinity of $\rho^-(770)$.

 \begin{figure}
 \includegraphics[width=\textwidth]{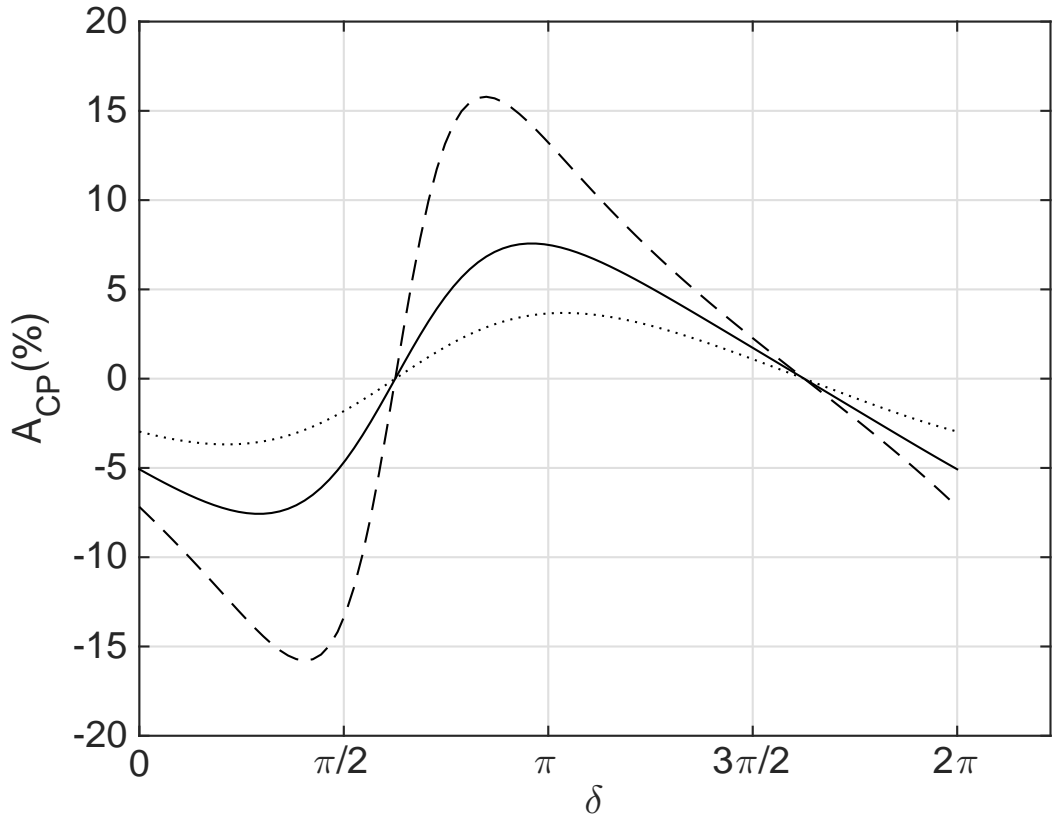}%
 \caption{\label{RegionalACPOL}$CP$ asymmetries in Region $\Omega_{\rm{OL}}$ as a function of the strong phase $\delta$. The dashed, solid, and dotted curves are for $F^{\Lambda_b\to p}/F^{\Lambda_b\to N^+}=0.5$, 1, and 2, respectively. }
 \end{figure}

In Fig. \ref{RegionalACPOL}, we present the regional $CP$ asymmetries in Region $\Omega_{\rm{OL}}$ as a function of the strong phase $\delta$.
The three curves in Fig. \ref{RegionalACPOL} correspond to the $CP$ asymmetries in Region $\Omega_{\rm{OL}}$ for $F^{\Lambda_b\to p}/F^{\Lambda_b\to N^+}=0.5$, 1, and 2, respectively, indicating that the regional $CP$ asymmetry is sensitive to the form factors.
From Fig. \ref{RegionalACPOL}, one can see that the interference of the decay amplitudes corresponding to the intermediate resonances $\rho^\mp(770)$ and $N^\pm(1440)$, together with proper strong phase $\delta$, does result in  the regional $CP$ asymmetry in the interference region of phase space.
Especially, the regional $CP$ asymmetry in Region $\Omega_{\rm{OL}}$ can reach as large as $\pm16\%$ when $F^{\Lambda_b\to p}/F^{\Lambda_b\to N^+}=0.5$.

To conclude, we want to point out that the interference of $\rho(770)$ with other baryonic resonances such as $N(1520)$ can also result in $CP$ asymmetries, given that the decay amplitudes corresponding to these baryonic resonances are comparable with that corresponding to $\rho(770)$ in the interference regions.
Besides, there can also be regional $CP$ asymmetries induced by the interference between the amplitudes corresponding to nearby baryonic resonances, for example, $N(1440)$ and $N(1520)$.
Similar behaviour to the latter one has already been proposed in $B$ meson decay processes \cite{Zhang:2013oqa} and observed by the LHCb collaboration \cite{Aaij:2013bla}.

%
\section*{Acknowledgments}
%
This work was partially supported by National Natural Science Foundation of China (Nos. 11175020, 11275025, 11447021, and 11575023), the construct program of the key discipline in Hunan province, and the Innovation Team of Nuclear and Particle Physics of USC.


\bibliographystyle{unsrtnat}
\bibliography{zzh}

%
\end{document}